# Calculation method of spin accumulations and spin signals in nanostructures using spin resistors

W. Savero Torres, A. Marty*, P. Laczkowski, L. Vila, M. Jamet and J-P. Attané

[1]Université Grenoble Alpes, INAC-SP2M, F-38000 Grenoble, France

[2]CEA, INAC-SP2M, F-38000 Grenoble, France

*alain.marty@cea.fr

**Abstract**

The understanding and calculation of spin transport are essential elements for the development of spintronics devices. Here, we propose a simple method to calculate analytically the spin accumulations, spin currents and magnetoresistances in complex systems. This can be used both for CPP experiments in multilayers and for multiterminal nanostructures made of semiconductors, oxides, metals and carbon allotropes.

## I.- Introduction

The development of spin-based devices relies on the injection and manipulation of spin currents in nanoscale systems. Therefore, the calculation and the understanding of spin transport at the nanoscale is essential to optimize spintronics devices.

In this context, several approaches to calculate spin currents and spin accumulations have been proposed in the last decades. The first analytical calculation of spin accumulation was provided by van Son et al.[1] in the case of a ferromagnet/normal metal interface. Valet and Fert[2] then derived the Boltzmann equation to provide the spin accumulation landscape in current perpendicular to the plane (CPP) experiments. This general model was rapidly extended to the case of semiconductors[3], providing the main conditions to optimize spin injection in such materials. Later on, Takahashi and Maekawa[4] used a similar approach to explain the spin transport in lateral spin valves, which was then refined by Hamrle et al. by including three dimensional effects[5,] and used by Kimura et al. to estimate the spin diffusion length by taking into account the spin sink in multi terminal devices.[6] It is worth mention that can also been calculated using matrix transfer approaches[7,8]

Spin-dependent diffusion equations are differential equations that can be solved step by step using boundary and continuity equations for charge and spin currents at each interface.[2] However, the calculation becomes quickly heavy in a complex structure comprising several elements. Here, we propose a simple and systematic method to derive mathematical expressions of the magnetoresistance in complex systems. It can be used both for CPP experiments in multilayers, and for multi-terminal nanostructures made of semiconductors, oxides, metals and carbon allotropes.

In the following, we will explain this method, and illustrate its simplicity by calculating the magnetoresistances of well-known systems. Then, we will show how it can be generalized and applied to systems possessing more complex geometries.

## II.- Method

This method primarily addresses the spin transport in multi-terminal spintronics structures, typically composed of several nanowires connected through transparent or non-transparent interfaces.[1,2,4,9,10] Such a system can be described by a network of spin resistors connected by nodes, where each node corresponds to a transparent interface. The concept of spin resistor is inspired from electric resistor concept, and concerns spin current transport.[10,11] Each wire and each non-transparent interface of the nanostructure (or each layer in the case of CPP transport) corresponds to a spin resistor (cf. Figs. 1a and 1b). As the objective is to describe the transport properties a current source is used to inject an electrical current between two nodes of the circuit, while a voltmeter detects the voltage between two nodes (not necessarily identical to the former).

An elementary wire is characterized by its material properties (electrical conductivity $\sigma$, spin polarization $\beta$, and spin diffusion length $\lambda$) and by its geometry (constant cross section of area A and length L). Within the wire, the flow of an electrical current $I_c$ and of a spin current $I_s$ take place due to gradient of electrochemical potentials (see cf. Fig 1a, b).

To provide an expression of the spin current along the wire, the electrochemical potentials $\mu(x)$ will be expressed hereafter in voltage units, and the spin dependent parameters will be denoted by the subscript $\pm$ (+ for spin up and - for spin down). Using this notation, the up and down current densities along the spin resistor can be expressed as[12] :

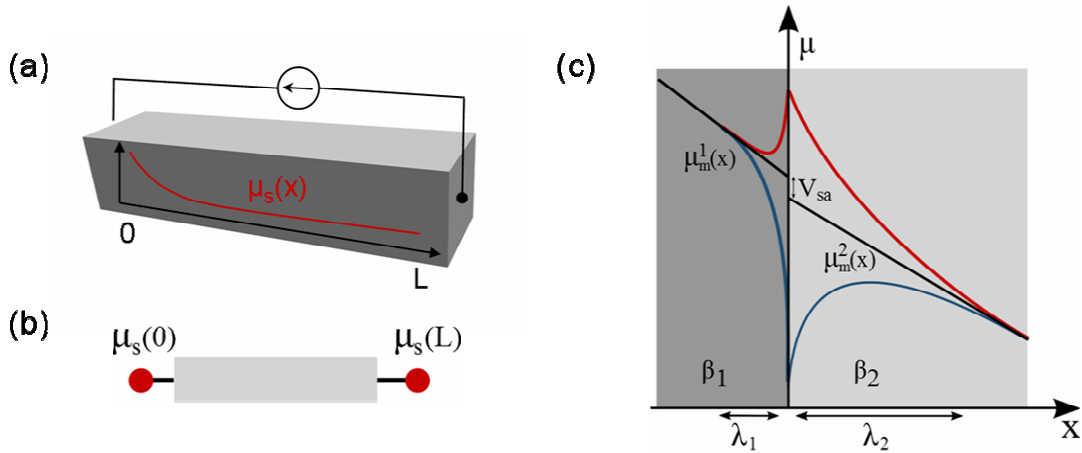

*Fig. 1. (a) Schematic representation of the spin accumulation along the wire for an electrical current applied along the x direction, and (b) symbol of the corresponding spin resistor. (c) Sketch showing the electrochemical potential landscape (in volt units) near the interface of two materials possessing polarisations $\beta_1$ and $\beta_2$ ($\beta_1 > \beta_2$). The diagonal black lines represent the electric potential in each material, which exhibits a drop at the interface ($V_{sa}$) produced by the spin accumulation. The red and blue curves represent the electrochemical potential of the majority and minority spin population in each material respectively.*

$$J_+ = -\sigma\frac{(1+\beta)}{2}\nabla\mu_+ \,;\, J_- = -\sigma\frac{(1-\beta)}{2}\nabla\mu_-$$

Where $\mu_\pm$ account for the electrochemical potentials for spin up and down populations, and where the spin polarization is defined as the asymmetry of the conductivities, $\beta = \frac{\sigma_+ - \sigma_-}{\sigma_+ + \sigma_-}$. The average potential $\mu_m$ and the spin accumulations $\mu_s$ are:

$$\mu_m = \frac{\mu_+ + \mu_-}{2} \,;\, \mu_s = \frac{\mu_+ - \mu_-}{2}$$

Then the charge and spin current density can be written as follows:

$$J_c = J_+ + J_- = -\sigma\nabla(\mu_m + \beta\mu_s) = -\sigma\nabla(V)\ldots(1)$$
$$J_s = J_+ - J_- = -\sigma(1-\beta^2)\nabla\mu_s + \beta J_c \ldots(2)$$

Eq. (1) shows that the charge current depends on both the average electrochemical potential and the spin accumulation through the gradient of a electric like potential $V(x)=\mu_m+\beta\mu_s$. While electrochemical potentials and therefore the average potential and spin accumulation are continuous at transparent interfaces, this electric like potential is not continuous. At the interface between two materials with different polarizations ($\beta_1$ and $\beta_2$), a potential drop between the materials appears due to the polarization change (cf. Fig 1c):

$$V_2 - V_1 = (\beta_2 - \beta_1)\mu_s$$

Additionally, Eq. (2) shows that the total spin current flowing along the wire is given by the sum of two contributions. The first contribution arises from the pure spin current, which is proportional to the gradient of the spin accumulation, while the second one arises from the spin polarized current. It is worth noting that in the case of non-magnetic materials the only contribution comes from the pure spin current, since there is no spin polarization ($\beta=0$).

Unlike charge current, spin current is not conservative due to spin flip processes. This leads to the differential equation governing the spin accumulation:[2]

$$\nabla^2\mu_s = \frac{\mu_s}{\lambda^2}\ldots(3)$$

Where, $\lambda$ is the spin diffusion length. With this second order linear differential equation, the spin accumulation along the wire can be expressed as a linear combination of the spin accumulations at the extremities:

$$\mu_s(x) = \frac{\sinh(\frac{L-x}{\lambda})}{\sinh(\frac{L}{\lambda})}\mu_s(0) + \frac{\sinh(\frac{x}{\lambda})}{\sinh(\frac{L}{\lambda})}\mu_s(L)\ldots(4)$$

Let us define the spin resistance R of the spin resistor as[1]:

$$R = \frac{\rho\lambda}{(1-\beta^2)A}\ldots(5)$$

Where $\rho$ is the usual electric resistivity of the wire.

Using Eqs. (2), (4) and (5), the spin current at the extremities of the spin resistor can be written as:

---

[1] Note that in the literature "spin resistance" sometimes denotes $\frac{\rho\lambda}{(1-\beta^2)A}$, sometimes $\frac{\rho\lambda}{A}$, with eventually the use of a star to distinguish ferromagnets from non-ferromagnets. For the sake of clarity we will in the following use the general definition of Eq. 5, with $\beta=0$ in normal metals.

$$I_s(0) = \frac{1}{R}\left(\frac{\mu_s(0)}{\tanh(\delta)} - \frac{\mu_s(L)}{\sinh(\delta)}\right) + \beta I_c \quad \ldots\ldots(6)$$

$$I_s(L) = \frac{1}{R}\left(\frac{\mu_s(0)}{\sinh(\delta)} - \frac{\mu_s(L)}{\tanh(\delta)}\right) + \beta I_c \quad \ldots\ldots(7)$$

Here, the dimensionless parameter, $\delta$, defined as $\delta = L/\lambda$, characterises the spin-flip ratio within the spin resistor. Unlike electric resistors, where charge current is constant along the wire, in spin resistors, the spin current is not the same at the two extremities. Eq. (6) and (7) provide the general expression for the spin current when the spin resistor is connected on both sides to other spin resistors, and when its length is comparable with its spin diffusion length (L~λ). However, such relations can be simplified when considering one of the three following cases.

1) When one extremity of the spin resistor, say (L) is unconnected, there is no charge current flowing within the wire, therefore the spin current at this unconnected extremity vanishes as well as the spin accumulation gradient : ∇μ$_s$=0. To satisfy this condition, the solution of Eq. (3) should be rewritten as:

$$\mu_s(x) = \frac{\cosh\left(\frac{L-x}{\lambda}\right)}{\cosh\left(\frac{L}{\lambda}\right)} \mu_s(0)$$

Then using Eq. (2) and (5) the spin current in that case is given by:

$$I_s(0) = \frac{\tanh(\delta)}{R} \mu_s(0)$$

2) When L>>λ (*i.e.* δ→∞), Eq. (6) is reduced to:

$$I_s(0) = \frac{\mu_s(0)}{R} + \beta I_c$$

Note that the spin current at one extremity does not depend on the spin accumulation at the other extremity. In experiments, such a condition is required for the long contacts probes.

3) When L<<λ, (*i.e.* δ→0) there is no spin flip in the wire and Eq. (6) can be written:

$$I_s(0) = \frac{1-\beta^2}{r}(\mu_s(0) - \mu_s(L)) + \beta I_c$$

where r=Rδ(1-β$^2$) is the electric resistance of the wire.

In the case of a non-transparent interface between two wires (for instance because of the presence of a tunnel barrier,[13,14,15,16] the interface has to be considered as a spin resistor connecting the two nodes, characterised by an electrical resistance r$_b$, a spin polarization coefficient $\gamma$ [17] and a spin flip ratio $\delta$. The spin resistance of the barrier is then defined as:

$$R_b = \frac{r_b}{(1-\gamma^2)\delta}$$

Fig.2 shows an example of a structure composed by several magnetic and non magnetic layers. As described above, it can be represented by a network of spin resistors connected by nodes (numbered red circles). The contacts probes for the current source and the voltmeter are not considered as nodes since they are far from the spin device and supposed to exhibit zero spin accumulation.

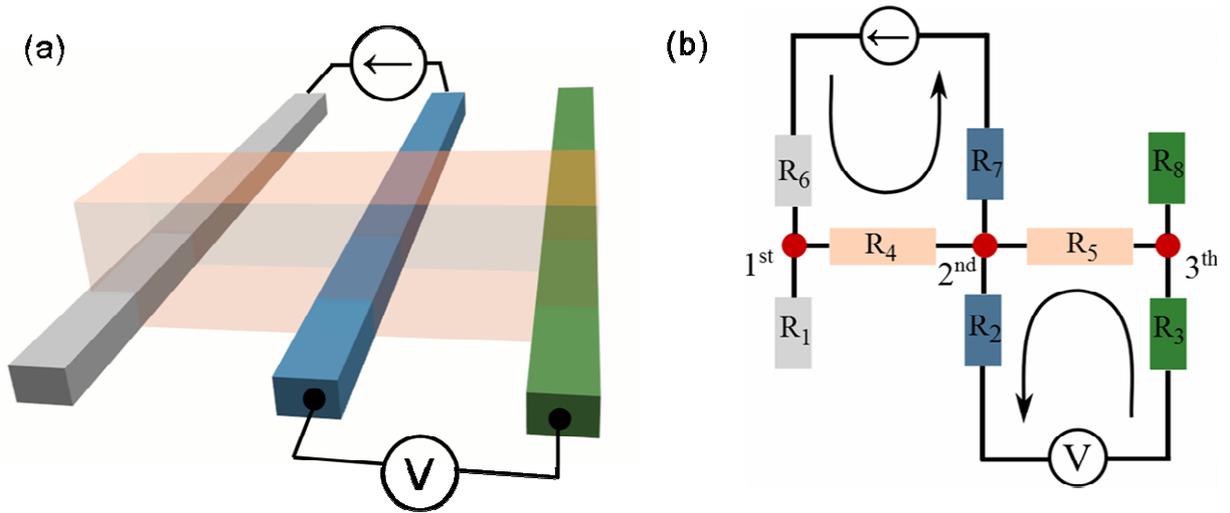

Fig2. (a) Schematic representation of a multiterminal nanostructure and (b) its equivalent spin resistor representation. The different colors correspond to different materials and the red circles are nodes characterizing transparent interfaces. The black arrows represent the current and voltage path along the structure.

To determine the spin accumulation at any node of this system, we consider the conservation of spin current at each node. Hence, the application of equations (6) and (7) to the system provide a set of linear equations that can be expressed simply using a matrix formulation of the form:

$$A\vec{\mu} = \vec{B}_{cur} I_c$$

Where $\vec{\mu}$ is a n-dimensional vector corresponding to the list of spin accumulation values at the n nodes. The matrix A and vector B are detailed below.

In a given probe configuration, the spin accumulation effects of the structure can be quantified through the evaluation of the spin accumulation resistance $R_{sa}=\Sigma V_{sa}/I_c$. $I_c$ is the charge current applied to the structure and $\Sigma V_{sa}$ is the sum of the voltage drops occurring at the nodes belonging to the loop including the voltmeter. Note that in non-local probe configurations the spin accumulation resistance is usually called the spin signal, whereas in local measurements it corresponds to the magnetoresistance. In most experiments, this spin accumulation resistance is measured as a function of an applied field, as it depends on the orientation of the magnetization in the ferromagnetic wires. In this approach, the spin accumulation resistance $R_{sa}$ for a magnetic state of the structure is given by the bilinear form :

$$R_{sa} = \vec{B}_{volt} A^{-1} \vec{B}_{cur} \quad ......(8)$$

In Eq. 8, $\vec{B}_{cur}$ is a n-dimensional vector whose elements depends on the path of the charge current. Its $i^{th}$ component can be determined as follow:

-if there is no charge current passing at the $i^{th}$ node, this component is equal to 0

-if a charge current flows from the resistor $R_p$ to the resistor $R_q$ through the $i^{th}$ node, the $i^{th}$ component of $B_{cur}$ is given by $\beta_p - \beta_q$

$\vec{B}_{volt}$ is the corresponding vector along the voltage path, i.e., the part of the circuit forming a loop with the

voltmeter. Its components can be determined as follows:

- If the $i^{th}$ node does not belong to the voltage path, the $i^{th}$ component of $\vec{B}_{volt}$ is 0

- If the voltage path goes from the resistor $R_p$ to the resistor $R_q$ through the $i^{th}$ node, the $i^{th}$ component of $\vec{B}_{volt}$ is given by $β_p$-$β_q$. Note that $\vec{B}_{cur}$ and $\vec{B}_{volt}$ depend on the magnetic state of the structure: $β_p$ is positive (resp. negative) if the magnetization of the ferromagnetic element $R_p$ points towards the direction corresponding to the up (resp. down) spins.

In addition, when the spin resistor corresponds to a tunnel barrier, the spin polarization β is usually labeled with the letter $\gamma$.

**A** is a nxn matrix characterizing the structure of the spin resistor lattice. The matrix elements $a_{ij}$ are constructed from equations. (6) and (7) in the following way:

**I.- Diagonal elements ($a_{ii}$):**

The **$a_{ii}$** element is derived by considering the $i^{th}$ node. It is equal to the sum of the following terms:

- $\dfrac{1}{R\tanh(\delta)}$ for each spin resistor connected to the $i^{th}$ node at one end and to another node at his other end (e.g., R4 and R5 in fig.2b)

- $\dfrac{1}{R}$ for each spin resistor connected to the $i^{th}$ node at one end and such as L>>λ. This is usually the case for long electrical contacts.

- $\dfrac{\tanh(\delta)}{R}$ for each spin resistor connected only to the to the $i^{th}$ node, the other end of the resistor being unconnected.

**II.- Non diagonal elements ($a_{ij}$):**

The matrix being symmetric by construction, only half of the non diagonal elements have to be calculated. The **$a_{ij}$** element is equal to :

- $-\dfrac{1}{R\sinh(\delta)}$ when the nodes *i* and *j* are connected through a single resistor (e.g., the nodes 1 and 2 in Fig 2b )

- 0 when the nodes *i* and *j* are not directly connected (e.g., the nodes *1* and 3 in Fig. 2b)

Finally, the spin signal amplitude Δ$R_{sa}$, which is the variation of the spin signal between two states of magnetization of the structure (I and II), is given by:

$$\Delta R_{sa} = R_{sa}^{I} - R_{sa}^{II}$$

**III.- Application**

In the following we will show how our method to calculate the spin signal amplitude Δ$R_{sa}$ can be applied to several well-known systems and more complex ones

**1.-F/N junctions**

**1a.- Junctions with transparent interfaces**

This structure, first studied by Johnson and Silsbee[18] and Van Son *et al.*[1], provides the simplest system where a spin accumulation creates a resistance contribution. It can be represented by two spin resistors, which are connected to one node at one extremity and to the electrical contacts at the other (see cf. Fig 3).

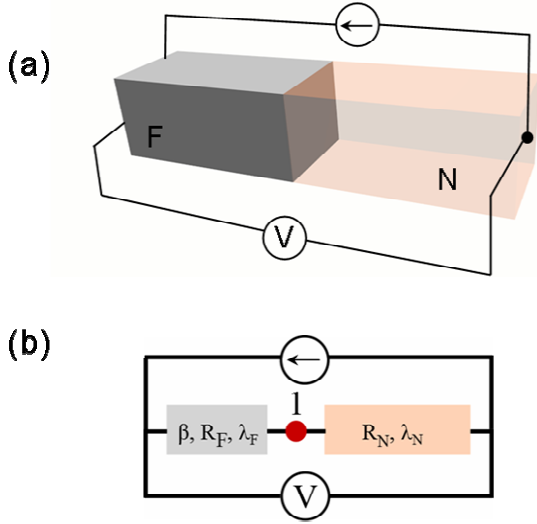

*Fig.3 (a) Schematic representation of a transparent F/N junction and (b) its equivalent spin resistor representation.*

According to the matrix construction method explained previously, as there is only one node the matrix A is a scalar whose value is given by:

$$A = \frac{1}{R_F} + \frac{1}{R_N}$$

As the current flow and voltage probe follow the same path, , $\vec{B}_{volt}$ and $\vec{B}_{cur}$ are equals: $B_{volt} = B_{cur} = \beta$. Therefore, the spin accumulation resistance $R_{sa}$ in this structure is simply given by:

$$R_{sa} = \vec{B}_{cur} A^{-1} \vec{B}_{volt} = \beta \left[ \frac{1}{R_F} + \frac{1}{R_N} \right]^{-1} \beta = \frac{\beta^2 R_F R_N}{R_F + R_N}$$

This is in agreement with the result obtained by Van Son *et al.*[1]

## 1b.- Junctions with inserted tunnel barriers (F/B/N)

The insertion of a tunnel barrier at a F/N interface is a simple system that is extensively used to increase the spin accumulation near the F/N interface,[13] leading to higher magnetoresistances and spin transfer torques,[19] and thus allowing to develop competitive memory and logic applications.[20,21]

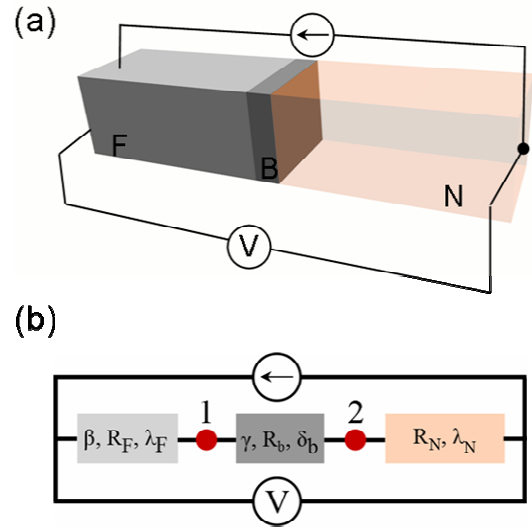

*Fig.4 Schematic representation of a F/B/N junction and its equivalent spin resistor representation*

The tunnel barrier is represented as a spin resistor, characterized by the spin flip parameters $\delta_b$, the polarization $\gamma$ and by the barrier spin resistance $R_b$ (cf. Fig. 4 a,b). As there are two nodes, **A** is a 2x2 matrix that can be constructed using the rules proposed previously:

$$A = \begin{pmatrix} \dfrac{1}{R_F} + \dfrac{1}{R_b \tanh\delta_b} & -\dfrac{1}{R_b \sinh\delta_b} \\ -\dfrac{1}{R_b \sinh\delta_b} & \dfrac{1}{R_N} + \dfrac{1}{R_b \tanh\delta_b} \end{pmatrix}$$

As the current and voltage follow the same path, the vectors $\vec{B}_{volt}$ and $\vec{B}_{cur}$ are also equals and given by:

$$\vec{B}_{volt} = \vec{B}_{cur} = \begin{pmatrix} \beta - \gamma \\ \gamma \end{pmatrix}$$

where β-γ and γ are the changes of polarization at the F/B and B/N interfaces, respectively. Therefore, after an algebraic reorganization, the spin accumulation resistance of this structure can be written as:

$$R_{sa} = \frac{\frac{(\beta-\gamma)^2}{R_N} + \frac{\gamma^2}{R_F} + \frac{\beta^2}{R_b \tanh(\delta_b)} - \frac{2\gamma(\beta-\gamma)\tanh(\delta_b/2)}{R_b}}{\frac{1}{R_b^2} + \left(\frac{1}{R_F} + \frac{1}{R_N}\right)\frac{1}{R_b \tanh(\delta_b)} + \frac{1}{R_F R_N}}$$

This expression can be simplified in the case of a barrier without spin flip ($\delta_b \to 0$). It takes the form:

$$R_{sa} = \frac{\gamma^2 R_N \frac{r_b}{1-\gamma^2} + (\beta-\gamma)^2 R_F \frac{r_b}{1-\gamma^2} + \beta^2 R_F R_N}{\frac{r_b}{1-\gamma^2} + R_F + R_N}$$

where $r_b$ is the electrical resistance of the barrier.

### 1c.-Double junction F/N/F

In this part, the method is applied to the case of double junctions with transparent F/N interfaces. This system, commonly known as a spin valve,[2,22] is the simplest multilayer device exhibiting Giant Magneto Resistance (GMR).[23]

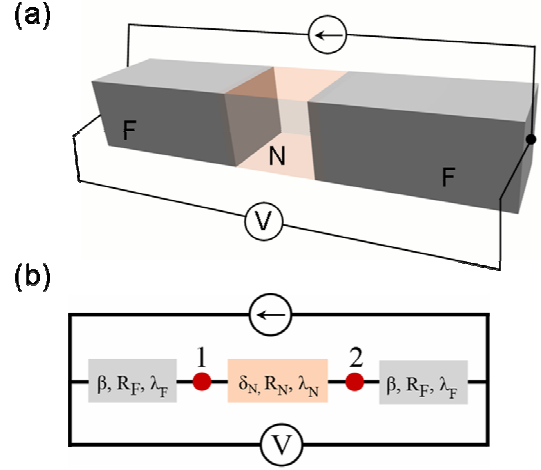

Fig.5 Schematic representation of a F/N/F junction and its equivalent spin resistor representation

To calculate the spin signal in this structure, let us take into account the spin resistor representation showed in Fig. 5. As the structure possesses two nodes, **A** is a 2x2 matrix with elements given by:

$$A = \begin{pmatrix} \frac{1}{R_F} + \frac{1}{R_N \tanh\delta_N} & -\frac{1}{R_N \sinh\delta_N} \\ -\frac{1}{R_N \sinh\delta_N} & \frac{1}{R_F} + \frac{1}{R_N \tanh\delta_N} \end{pmatrix}$$

The current and voltage probe follow the same path, therefore when both magnetizations are parallel the associated vectors are given by:

$$\vec{B}_{volt} = \vec{B}_{cur} = \begin{pmatrix} \beta \\ -\beta \end{pmatrix}$$

Whereas, when they are antiparallel:

$$\vec{B}_{volt} = \vec{B}_{cur} = \begin{pmatrix} \beta \\ \beta \end{pmatrix}$$

The spin signal amplitude is $\Delta R_s = R_{sa}^{AP} - R_{sa}^{P}$, where $R_{sa}^{P}$ and $R_{sa}^{AP}$ are the spin accumulation resistances in the parallel and antiparallel states, and are calculated using Eq. 8.

It leads to

$$\Delta R_s = \frac{8\beta^2 R_F^2 R_N}{(R_N+R_F)^2 e^{\delta} - (R_N-R_F)^2 e^{-\delta}}$$

which is equivalent to the expression obtained in Ref 2 and 7.

## 2.- Lateral spin valves

Let us consider a lateral spin valve,[9,18] with inserted tunnel junctions at the interfaces, in order to calculate the analytical expression of the spin signal (cf. Fig. 6).

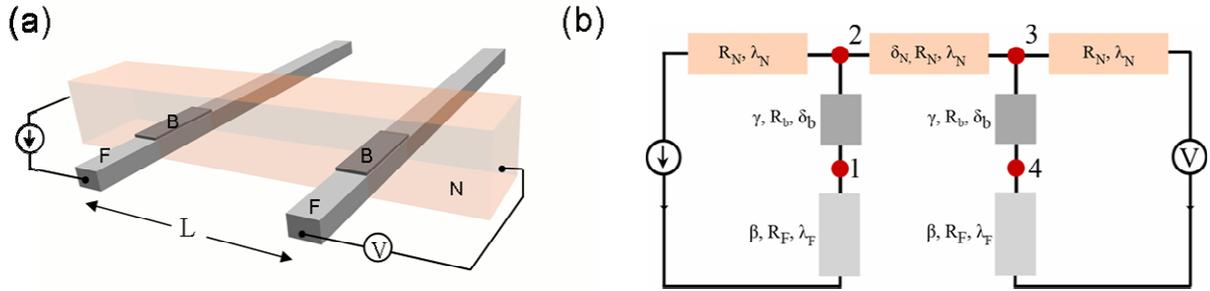

*Fig. 7 (a) Schematic representation of a lateral spin valve with inserted tunnel junctions and (b) the equivalent spin resistor representation*

Following the proposed construction rules, The **A** matrix is:

$$A = \begin{pmatrix} \frac{1}{R_F}+\frac{1}{R_b\tanh\delta_b} & -\frac{1}{R_b\sinh\delta_b} & 0 & 0 \\ -\frac{1}{R_b\sinh\delta_b} & \frac{1}{R_N}+\frac{1}{R_b\tanh\delta_b}+\frac{1}{R_N\tanh\delta_N} & -\frac{1}{R_N\sinh\delta_N} & 0 \\ 0 & -\frac{1}{R_N\sinh\delta_N} & \frac{1}{R_N}+\frac{1}{R_b\tanh\delta_b}+\frac{1}{R_N\tanh\delta_N} & -\frac{1}{R_b\sinh\delta_b} \\ 0 & 0 & -\frac{1}{R_b\sinh\delta_b} & \frac{1}{R_F}+\frac{1}{R_b\tanh\delta_b} \end{pmatrix}$$

Here, the current and the voltage follow different paths, and therefore the vectors $B_{cur}$ and $B_{volt}$ are different. When the magnetizations of the electrodes are parallel, these vectors are given by:

$$\vec{B}_{cur} = \begin{pmatrix} \beta-\gamma \\ \gamma \\ 0 \\ 0 \end{pmatrix}; \vec{B}_{volt} = \begin{pmatrix} 0 \\ 0 \\ -\gamma \\ \gamma-\beta \end{pmatrix}$$

As the current and voltage paths involve only two nodes (see Fig 5b), their corresponding vectors possess only two non vanishing elements. Here again, the spin signal amplitude is determined from: $\Delta R_s = R_{sa}^P - R_{sa}^{AP}$ where the variation of the magnetic state is taken into account by changing the spin polarisations from β→-β in the layers and from γ→-γ in the barriers.

The complete analytical expression can be obtained by applying Eq. 8 giving a rather large expression:

$$\Delta R_{sa} = \frac{4R_N R_b^2 \left((\beta-\gamma)R_F + \gamma R_b sh(\delta_b) + \gamma R_F ch(\delta_b)\right)^2}{e^{\delta_N}\left(ch(\delta_b)R_b(R_N+2R_F)+sh(\delta_b)(R_N R_F + 2R_b^2)\right)^2 - e^{-\delta_N}R_N^2\left(ch(\delta_b)R_b + sh(\delta_b)R_F\right)^2}$$

If we restrict our analysis to the case where there is no spin flip at the interfaces (*i.e.* δb→0), the spin signal amplitude is given by:

$$\Delta R_{sa} = \frac{4R_N(\gamma \frac{r_b}{1-\gamma^2} + \beta R_F)^2}{\left(R_N + 2\frac{r_b}{1-\gamma^2} + 2R_F\right)^2 e^{\delta_N} - R_N^2 e^{-\delta_N}}$$

Now, by considering the limit cases, we can easily obtain the following results:

$$\Delta R_s = R_N \gamma^2 Exp(-\delta_N)$$

when $r_b \gg R_F$ and $R_N$

$$\Delta R_s = \frac{2R_N(\gamma \frac{r_b}{1-\gamma^2})^2}{R_N^2 sh(\delta_N)}$$

when $R_F \ll r_b \ll R_N$

$$\Delta R_s = \frac{2(\beta R_F)^2}{R_N sh(\delta_N)}$$

when $r_b \ll R_F \ll R_N$

The expressions obtained in the simplified cases are in agreement with those obtained by Takahashi et Maekawa[4], and have been used in experimental systems to characterize the spin-dependent properties in non magnetic materials[5,9,24,25].

### 3.- More complex geometries

Up to now we have studied simple and well known systems, in order to check the benefit of this approach, and to underline the facility of the proposed method to derive mathematical expression. In the following, we will show how it can be used in more complex cases, considering as an example a structure with four ferromagnetic wires with inserted tunnel junctions.

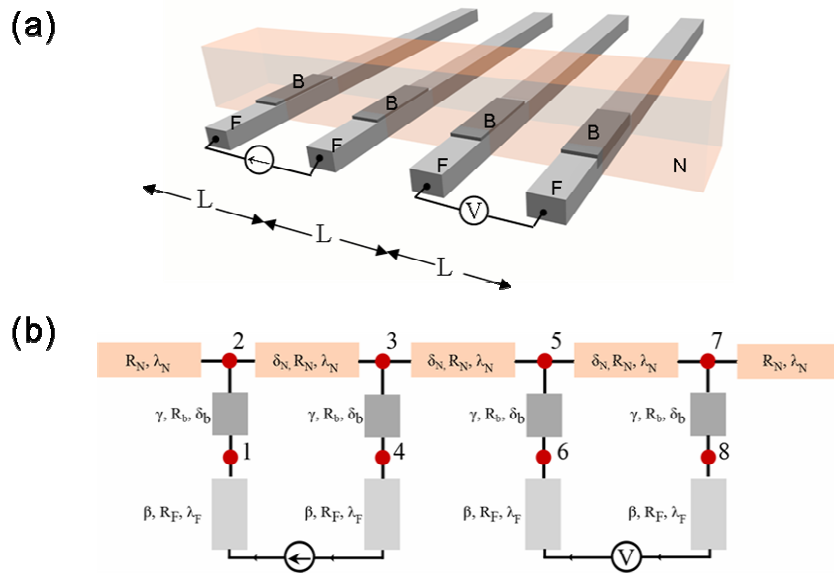

*Fig. 7 (a) Schematic representation of a complex geometry composed by several ferromagnetic wires and tunnel junctions. (b) Sketch showing the spin resistor representation.*

In that case all the ferromagnetic elements and all the tunnel junctions are supposed to be identical. The nodes having similar surroundings can be gathered into three groups; {2,7}, {3,5} and {1,4,6,8}. Symmetries of this geometry thus imply that the 8x8 matrix **A** can be written using only five independent coefficients:

$$A = \begin{pmatrix} a_{11} & a_{12} & a_{13} & 0 & 0 & 0 & 0 & 0 \\ a_{12} & a_{22} & 0 & 0 & 0 & 0 & 0 & 0 \\ a_{13} & 0 & a_{33} & a_{12} & a_{13} & 0 & 0 & 0 \\ 0 & 0 & a_{12} & a_{22} & 0 & 0 & 0 & 0 \\ 0 & 0 & a_{13} & 0 & a_{33} & a_{12} & a_{13} & 0 \\ 0 & 0 & 0 & 0 & a_{12} & a_{22} & 0 & 0 \\ 0 & 0 & 0 & 0 & a_{13} & 0 & a_{11} & a_{12} \\ 0 & 0 & 0 & 0 & 0 & 0 & a_{12} & a_{22} \end{pmatrix}$$

With:

$$a_{11} = \frac{1}{R_N} + \frac{1}{R_N \tanh\delta_N} + \frac{1}{R_b \tanh\delta_b}; a_{22} = \frac{1}{R_F} + \frac{1}{R_b \tanh\delta_b};$$

$$a_{33} = \frac{2}{R_N \tanh\delta_N} + \frac{1}{R_b \tanh\delta_b}; a_{12} = -\frac{1}{R_b \sinh\delta_b};$$

$$a_{13} = -\frac{1}{R_N \sinh\delta_N}$$

Then, to write $\vec{B}_{volt}$ and $\vec{B}_{cur}$, one has to consider the current and voltage path showed in Fig. 7. If the four ferromagnetic electrodes possess parallel magnetizations, $\vec{B}_{cur}$ and $\vec{B}_{volt}$ can be written as:

$$B_{cur} = \begin{pmatrix} \gamma \\ \beta - \gamma \\ -\gamma \\ \gamma - \beta \\ 0 \\ 0 \\ 0 \\ 0 \end{pmatrix}, B_{volt} = \begin{pmatrix} 0 \\ 0 \\ 0 \\ 0 \\ \gamma \\ \beta - \gamma \\ -\gamma \\ \gamma - \beta \end{pmatrix}$$

When considering other magnetic configurations of the electrodes, $\vec{B}_{cur}$ and $\vec{B}_{volt}$ can constructed by replacing β by -β and ɣ by –ɣ for the electrode with reversed magnetization. Note that in this article we focused on the calculation of the spin accumulation resistance given by Eq. (8), as this resistance is usually the most important parameter in experiments. Now, let us note $\vec{\mu}_S$ the vector whose $i^{th}$ component is the spin accumulation at the $i^{th}$ node. The knowledge of $\vec{\mu}_S$ is obtained by using: $\vec{\mu}_S = I_C A^{-1} \vec{B}_{cur}$, and the whole spin accumulation landscape can then be calculated using Eq. (4).

## IV.- Conclusion

The proposed method allows thus to obtain an analytical expression of the spin accumulation and of the spin signal in any complex structure. Still, there are three limitations to this method. The first one is that of the Valet-Fert model: It applies only to systems with collinear magnetization (*i.e.,* spins can only be up or down). The second one is common to all existing analytical resolutions of the Valet-Fert differential equations: the calculation is one-dimensional, in the sense that the electrochemical potentials depends only of the position along the length of the wire. The last limitation comes from our will to deal with spin accumulation alone: in this model the charge current must be known prior to calculations. This is possible if the imposed current follows a unique path, in the sense that there is no closed loop in the circuit. Note this situation is encountered in all proposed devices so far.

To sum up, we proposed a method to calculate relatively simply the spin signal in both CPP measurements and multi-terminal nanostructures. This method has been applied to well-known systems, and we demonstrated how this method could be applied to more complex geometries.